\documentclass[lettersize,journal]{IEEEtran}
\usepackage{amsmath,amsfonts, amsthm}
\usepackage{algorithmic}
\usepackage{algorithm}
\usepackage{array}
\usepackage{textcomp}
\usepackage{stfloats}
\usepackage{url}
\usepackage{verbatim}
\usepackage{graphicx}
\usepackage{cite}
\usepackage{physics}
\usepackage[utf8]{inputenc}
\usepackage{fontenc}
\usepackage[table]{xcolor}
\definecolor{first}{HTML}{DE2014}
\definecolor{second}{HTML}{DE8514}
\usepackage{subcaption}
\usepackage{booktabs}  
\usepackage{algorithm}
\usepackage{algorithmic}
\usepackage[colorlinks=true, linkcolor=blue, citecolor=blue, urlcolor=blue]{hyperref}
\newcommand{\thetavec}{\pmb{\theta}}
\usepackage{subfloat}
\usepackage{multirow}
\usepackage{float}
\usepackage{aliascnt}

\newtheorem{theorem}{Theorem}[section]

\newaliascnt{proposition}{theorem}
\newtheorem{proposition}[proposition]{Proposition}
\aliascntresetthe{proposition}

\newtheorem{definition}[theorem]{Definition}

\usepackage[textsize=small]{todonotes}

\newcommand\QP{\text{QP}}

\def\UGC{\mathbf{U}}

\def\pri{\mathbf{x}}
\def\ugc{\mathbf{u}}
\def\dr{\mathbf{z}}

\def\encugc{\hat{\mathbf{u}}}
\def\ris{R^{(k)}_{s}}
\def\rit{R^{(k)}_{t}}
\def\dis{D^{(k)}_{s}}
\def\dit{D^{(k)}_{t}}
\def\lams{\lambda_{s}^{(k)}}
\def\lamt{\lambda_{t}^{(k)}}
\def\OursD{DSD}
\def\OursRD{RDSD}
\def\quasat{QS}
\def\dissat{DS}
\def\primse{P-MSE}
\def\armse{D-MSE}
\def\inmse{I-MSE}
\def\idmse{ID-MSE}

\def\thetavec{\pmb{\theta}}

\DeclareMathOperator*{\argmin}{arg\,min}
\newcommand{\img}{\mathbf{x}}
\newcommand{\cimg}{\hat{\mathbf{x}}}

\newcommand{\nummb}{M}
\newcommand{\quantstep}{q}

\begin{document}

\title{Avoiding Quality Saturation in UGC Compression Using Denoised References}

\author{Xin Xiong, Samuel Fern\'andez-Mendui\~na, Eduardo Pavez, Antonio Ortega, Neil Birkbeck, Balu Adsumilli

\thanks{X. Xiong, S.  Fern\'andez-Mendui\~na, E. Pavez and A. Ortega are with the University of Southern California (email: \{xiongxin, samuelf9, pavezcar, aortega\}@usc.edu). N. Birkbeck and B. Adsumilli are with Youtube/Google (email: \{birkbeck, badsumilli\}@google.com). This work was funded in part by a gift from YouTube.}
}

\markboth{Journal of \LaTeX\ Class Files,~Vol.~14, No.~8, August~2021}%
{Shell \MakeLowercase{\textit{et al.}}: A Sample Article Using IEEEtran.cls for IEEE Journals}


\maketitle

\begin{abstract}
Video-sharing platforms must re-encode large volumes of noisy user-generated content (UGC) to meet streaming demands. 
However, conventional codecs, which aim to minimize the mean squared error (MSE) between the compressed and input videos, can cause quality saturation (\quasat{}) when applied to UGC, i.e., increasing the bitrate preserves input artifacts without improving visual quality.
A direct approach to solve this problem is to detect \quasat{} by repeatedly evaluating a non-reference metric (NRM) on videos compressed with multiple codec parameters, which is inefficient.
In this paper, we re-frame UGC compression and \quasat{} detection from the lens of noisy source coding theory: rather than using a NRM, we compute the MSE with respect to the denoised UGC, which serves as an alternative reference (\armse{}). 
Unlike MSE measured between the UGC input and the compressed UGC, \armse{} saturates at non-zero values as bitrates increase, a phenomenon we term distortion saturation (\dissat{}). 
Since \armse{} can be computed at the block level in the transform domain, we can efficiently detect \dissat{} without coding and decoding with various parameters. 
We propose two methods for \dissat{} detection: distortion saturation detection (\OursD{}), which relies on an input-dependent threshold derived from the \armse{} of the input UGC, and rate-distortion saturation detection (\OursRD{}), which estimates the Lagrangian at the saturation point using a low-complexity compression method. 
Both methods work as a pre-processing step that can help standard-compliant codecs avoid \quasat{} in UGC compression.
Experiments with AVC show that preventing encoding in the saturation region, i.e., avoiding encoding at QPs that result in \quasat{} according to our methods, achieves BD-rate savings of $8\%$-$20\%$ across multiple different NRMs, compared to a naïve baseline that encodes at the given input \QP\ while ignoring \quasat{}.

\end{abstract}

\section{Introduction}
\IEEEPARstart{U}{ser}-generated content (UGC) \cite{wang2019youtube}, or non-professional video, has become central to video-sharing platforms such as YouTube and TikTok.
These videos are often noisy due to amateur production and subpar equipment. Since all uploaded videos have been compressed (at least once), prior compression also results in artifacts. 
Once uploaded, service providers \textit{re-encode} UGC at multiple resolutions and quality levels for video streaming \cite{seufert2014survey}. 
However, since most codecs are designed to minimize the mean squared error (MSE) between the input and compressed videos, increases in bitrate may not improve perceived video quality and instead will result in preserving artifacts present in the original UGC input, an effect we call \textit{quality saturation (QS)} \cite{pavez2022compression, xiong2023rate, wang2019youtube}.
While \quasat\ could be detected by quantifying the perceptual quality of UGC inputs compressed at different bitrates, e.g., by  finding Mean Opinion Scores (MOS) \cite{wang2012perceived},  such a large‑scale perceptual evaluation is practically infeasible.

\begin{figure*}[!t]
\centering
\includegraphics[width=1\linewidth]{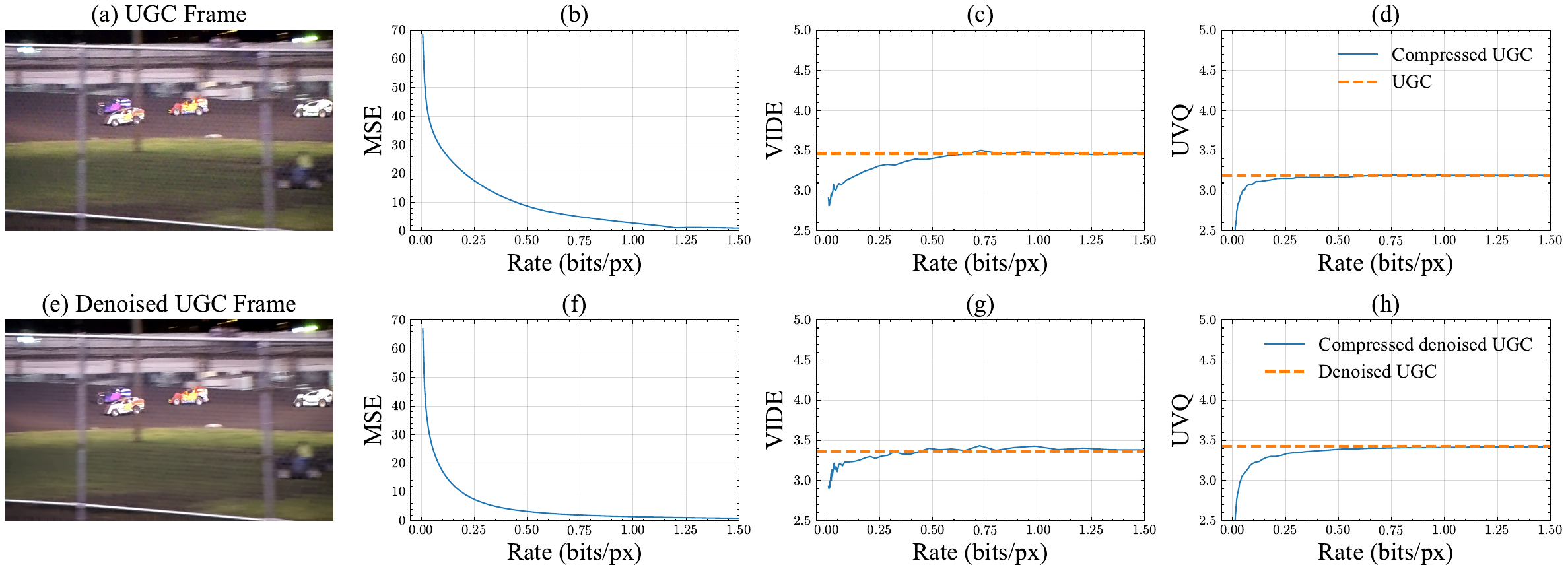}
\caption{(a) A frame from a UGC video; (b–d) rate-distortion (RD) curves from compressing the UGC video with AVC. VIDE \cite{tu2021ugc} and UVQ \cite{wang2021rich} are NRMs that predict visual quality. While MSE (b) converges to zero distortion (i.e., perfect approximation to the input) as bitrate increases, both VIDE (c) and UVQ (d) do not reach perfect quality at high bitrates. (e) A frame from the denoised UGC video; (f-h) RD curves from compressing this denoised UGC video with AVC. Similar to (c-d), VIDE (g) and UVQ (h) saturate at high bitrates when encoding the denoised UGC.}
\label{fig:I1}
\end{figure*}

As an alternative to MOS-based \quasat ~detection, non-reference metrics (NRMs) \cite{wu2023exploring, tu2021ugc, wang2021rich, tu2021rapique}, which predict perceptual quality without a reference, can also reveal \quasat.
As shown in \autoref{fig:I1}(c) and \autoref{fig:I1}(d), the perceptual quality of the compressed videos, as measured by NRMs, saturates at the input quality as the rate increases, while the MSE continues to decrease towards zero (\autoref{fig:I1}(b)). 
The example in \autoref{fig:I1} suggests that conventional compression systems are inefficient in UGC compression: \textit{increasing the bitrate leads to perfect reconstruction (zero MSE) of a noisy reference and thus preserves artifacts that do not enhance perceptual quality.} 
While NRMs provide a straightforward approach to detect and avoid \quasat{} by analyzing the rate-quality curve, this approach requires multiple rounds of encoding, decoding, and metric evaluation to generate the rate-quality points, which is computationally impractical.

In this paper, following \cite{pavez2022compression, xiong2023rate}, we address the problem of, \emph{given a UGC signal and a codec, 
choosing coding parameters to maximize quality while avoiding allocating resources to encode artifacts and noise.}
We aim to avoid \quasat{}  without resorting to costly computation of NRMs for each candidate coding choice 
(as in \autoref{fig:I1}(c) and \autoref{fig:I1}(d)).
We do not consider postprocessing methods, e.g., denoising \cite{zhang2017beyond, zhang2023practical,brummer2023importance} and restoration \cite{li2024promptcir}. These can improve quality but do not prevent \quasat{}: since they are applied after decoding, and excessive bitrate may have already been used to represent accurately the noise. 

Closest to our proposed work are pre-processing methods that classify UGC inputs based on statistical properties \cite{john2020rate, ling2020towards, feng2024content} or perceptual quality \cite{wang2020video}, 
and use preset codec parameters tailored to each UGC class. 
However, these methods do not explicitly model or account for \quasat{} when deriving their presets and thus they cannot avoid the occurrence of \quasat{}. 

\begin{figure}[!t]
\centering
\includegraphics[width=1\linewidth]{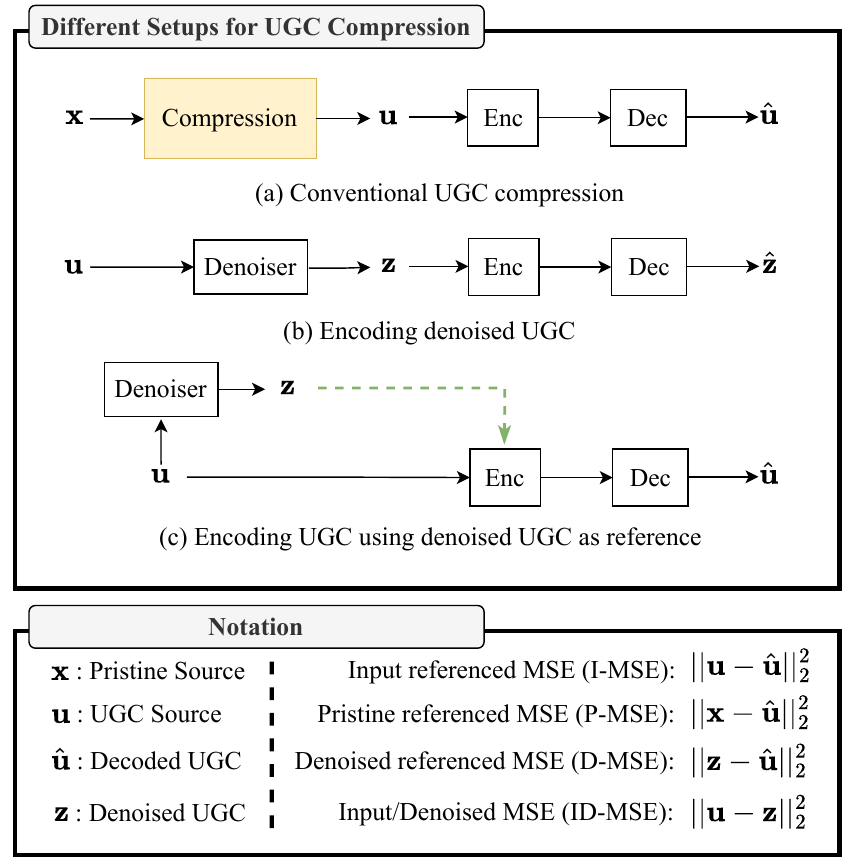}
\caption{Different encoding setups for UGC compression.
}
\label{fig:Sys_Compare}
\end{figure}

\begin{figure*}[!t]
\centering
\includegraphics[width=1\linewidth]{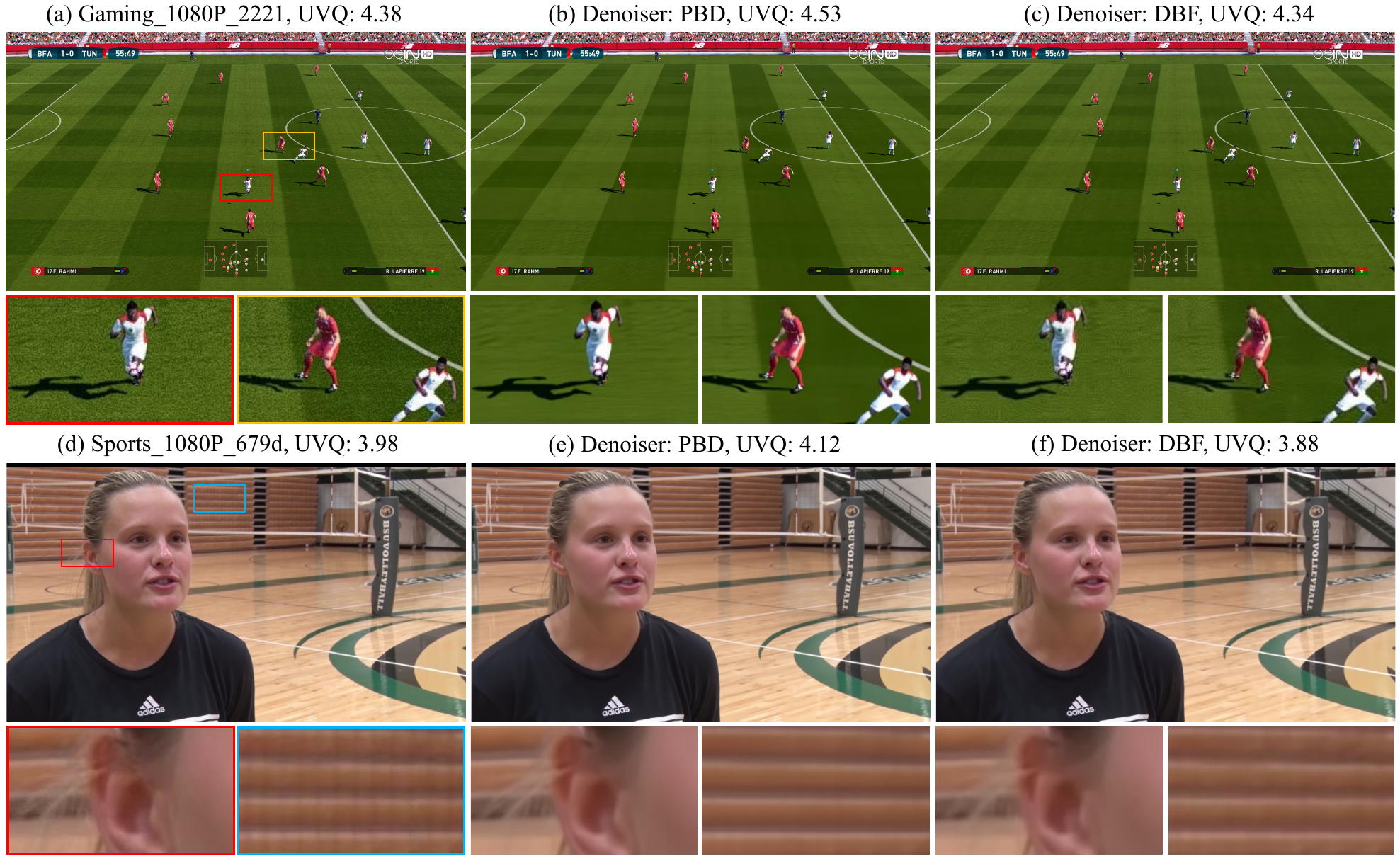}
\caption{
Examples of UGC frames and their denoised version. From left to right: original UGC frame, and denoised UGC frames using Practical Blind Denoiser (PBD) \cite{zhang2023practical} and FFmpeg's De-Blocking Filter (DBF) \cite{tomar2006converting}. Both denoisers over-smooth the grass (top) and remove the net (bottom), demonstrating that current denoisers can alter the image content. Moreover, a suboptimal denoiser (e.g., DBF) can produce lower-quality results than the original UGC, as measured by non-reference metrics.
}
\label{fig:Exp_Denoiser}
\end{figure*}

In our previous work \cite{pavez2022compression} we formulated UGC compression as a noisy source coding problem \cite{berger1971rate,dobru1962source,wolf1970transmission,al1998lossy,ephraim1988unified,fischer1990estimation}, where the UGC video $\ugc$ and the (unknown) pristine video $\pri$ are the noisy and clean sources, respectively. 
In this framework, the ideal distortion metric is the MSE with respect to the unavailable pristine reference $\pri$, (\primse{}: $\Vert\pri -\hat{\ugc}\Vert_2^2$, \autoref{fig:Sys_Compare}), which cannot be assumed to converge to zero (see \autoref{Method_A}). 
We call this phenomenon distortion saturation (\dissat) \cite{pavez2022compression}, which is similar to \quasat{} in NRMs. According to noisy source coding theory, optimality can be achieved by optimally encoding the optimally denoised input \cite{pavez2022compression}, which suggests a UGC coding strategy based on denoising followed by encoding (\autoref{fig:Sys_Compare} (b)). 
However, a practical implementation of this system using off-the-shelf denoisers may be suboptimal \cite{yamashita2002relative}
and produce undesirable changes in the video.  
For example, \autoref{fig:Exp_Denoiser} shows that denoisers can alter UGC content and oversmooth certain regions \cite{blau2019rethinking}.
These limitations in available denoisers are problematic for two reasons. 
First, in some cases (see \autoref{fig:I1} (e-h)) \quasat\ persists when encoding the denoised UGC \cite{xiong2023rate}, 
because the encoder approximates as well as possible (given the target rate) the denoised UGC whose quality in general does not match that of the pristine source. 
Second, some users may prefer not to denoise before encoding, since it can significantly change the appearance of the content they upload (see \autoref{fig:Exp_Denoiser}).  

Hence, in this paper, we propose to directly encode the original UGC while \textit{using the denoised UGC only as an alternative reference for distortion computation} \cite{pavez2022compression,xiong2023rate}. We consider this to be a design choice that users may favor depending on the reliability of available denoisers.  

\begin{figure}[!t]
\centering
\includegraphics[width=1\linewidth]{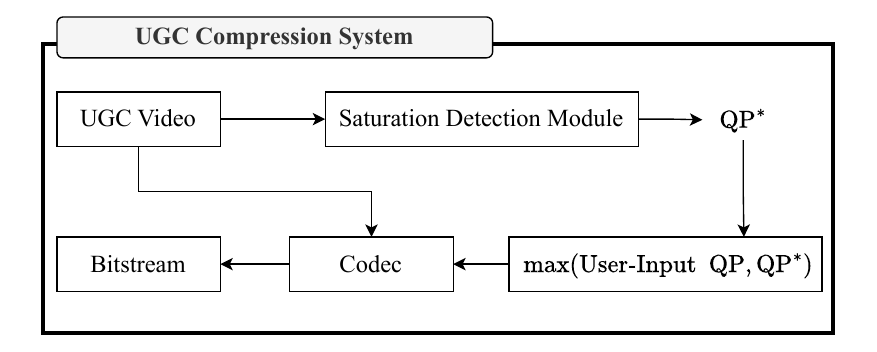}
\caption{The proposed UGC compression system. Given a UGC video, our saturation detection method produces a saturation quality parameter $\QP^*$, indicating where \dissat\ occurs. Ideally, if $\QP^*$ exactly matches where \quasat\ happens, encoding the video with any $\QP$ lower than $\QP^*$ does not improve quality but wastes bitrate. Therefore, if a user specifies a desired $\QP$ for encoding, we send the larger value between the user-input $\QP$ and $\QP^*$ to the codec.}
\label{fig:I3_System}
\end{figure}

Our proposed method follows the setup shown in \autoref{fig:Sys_Compare} (c), where we use the denoised UGC $\dr$ only as side information to encode the UGC input $\ugc$.  In order for our UGC compression system to be compatible with existing video coding pipelines, we perform DS detection as a pre-processing step, and then communicate to the video encoder a range of encoding parameters (e.g., $\QP$ values) that avoid \dissat{}. This system is depicted in \autoref{fig:I3_System}. The saturation detection module from \autoref{fig:I3_System} computes three distortion functions comparing $\ugc$, $\dr$ and $\hat{\ugc}$. 
The conventional MSE (\inmse{}: $\Vert\ugc -\hat{\ugc}\Vert_2^2$) approaches zero as the rate increases (\autoref{fig:I1} (b)). Since the MSE with respect to the pristine reference, (\primse{}: $\Vert\pri -\hat{\ugc}\Vert_2^2$) is never available, given that $\dr$ is an estimate of $\pri$, we consider MSE with respect to the denoised reference (\armse\: $\Vert \dr - \encugc \Vert^2_2$), which  saturates to a non-zero value (similar to \primse).  We also compute the MSE between the input UGC $\ugc$ and the denoised UGC $\dr$ (\idmse{}: $\Vert \ugc-\dr \Vert_2^2$), which serves as a noise estimate. These three distortions are used by the saturation detection module to detect \dissat{} in \armse. 
Our experiments show that avoiding \dissat\ in \armse\ also helps prevent \quasat\ \cite{xiong2023rate}.
Compared to detecting \quasat{} with NRMs, \dissat\ detection is much more efficient because it can be computed block-wise in the transform domain (due to Parseval's identity), without decoding the image.

Although we theoretically demonstrated the existence of \dissat\ in our initial work \cite{pavez2022compression}, this approach was highly sensitive to the choice of denoiser and was applied only to synthetic UGC images \cite{xiong2023rate} (see \autoref{Method_B}). 
Our subsequent work \cite{xiong2023rate} proposed a UGC compression system (see \autoref{fig:I3_System}) that uses \dissat{} to prevent \quasat{}. However, \cite{xiong2023rate} relies on an ad-hoc \dissat\ detection criterion and requires manually selecting a codec-dependent threshold, which is difficult to determine empirically.
Our experiments show that our new approach significantly outperforms \cite{xiong2023rate} on real UGC videos (see \autoref{Exp_D}).

This work improves our prior studies \cite{pavez2022compression, xiong2023rate} in two ways. 
1) \emph{Methodologically}, building on the noisy source coding formulation from \cite{pavez2022compression}, we propose new  \dissat\ detection methods that significantly outperform \cite{xiong2023rate} in both complexity and \quasat{} prevention.
The proposed methods, distortion-based saturation detection (\OursD) and rate-distortion-based saturation detection (\OursRD), are used before encoding, do not modify the UGC input fed to the encoder, and are compatible with standard-compliant codecs.
2) \emph{Empirically}, we perform an \textit{extensive evaluation on a large, real‑world UGC video dataset\cite{wang2019youtube}}, rather than synthetic UGC clips or images as in \cite{pavez2022compression,xiong2023rate}. Furthermore, we provide quantitative results to validate the performance of our methods in UGC compression.

\OursD\ compares the expected quantization error to a threshold given by the MSE between the input UGC and its denoised version (\idmse: $\Vert \dr - \ugc \Vert^2_2$), which provides an easy-to-compute estimate of the noise intensity (see \autoref{Method_B}).
Compared to \cite{pavez2022compression}, using this threshold is more robust to the choice of denoiser and performs well on real‑world UGC videos.
This threshold is content-adaptive, codec-independent, and can be selected automatically, unlike the one proposed in \cite{xiong2023rate}, which had to be selected manually. 
Furthermore, \dissat\ detection can be achieved by replacing distortion computation at each quantization step with its expectation, avoiding exhaustively searching RD points.
Since most modern video codecs control the RD trade-off using a Lagrange multiplier $\lambda$ \cite{ortega1998rate},  \OursRD\ estimates $\lambda$ at the saturation point identified by \OursD\ (see \autoref{Method_C}).  Similar to \cite{xiong2023rate}, we generate RD points by using a \textit{low-complexity codec}, e.g., a simplified version of  AVC \cite{richardson2011h} (See \autoref{estimate_lambda}).
Furthermore, we show it is possible to transfer the resulting saturation Lagrange multiplier $\lambda^{*}_s$ from a source codec, e.g., a low-complexity codec, to a target codec. 
In the high-rate regime, when the low-complexity codec and the target codec achieve the same distortion, i.e., the onset of \dissat, we prove that their corresponding Lagrange multipliers are proportional and propose a method to estimate this proportionality constant (see \autoref{transfer_lambda}).

We evaluate our proposed UGC compression system on the YouTube UGC dataset~\cite{wang2019youtube}, using UVQ~\cite{wang2021rich}, SimpleVQA~\cite{sun2022deep}, DOVER~\cite{wu2023dover}, and FasterVQA~\cite{wu2022fasterquality} as NRMs.
As a baseline, we consider a simple approach in which UGC clips are encoded at various \QP\ values without accounting for \quasat{}. 
By modifying this baseline to avoid encoding at \QP\ values that our methods identify as producing \quasat{}, we obtain up to $15\%$ bitrate savings with minimal UVQ degradation  
(\autoref{Exp_D}). 
Our methods are agnostic to the choice of denoiser and substantially outperform the approach in~\cite{xiong2023rate}. While \textit{our methods do not specifically optimize for UVQ or any other NRMs}, 
they achieve BD-rate savings of $8\%-20\%$ (\autoref{Exp_Results_Rate_Quality}) across multiple NRMs relative to the naive baseline.

\begin{figure*}[htbp]
\centering
\includegraphics[width=1\linewidth]{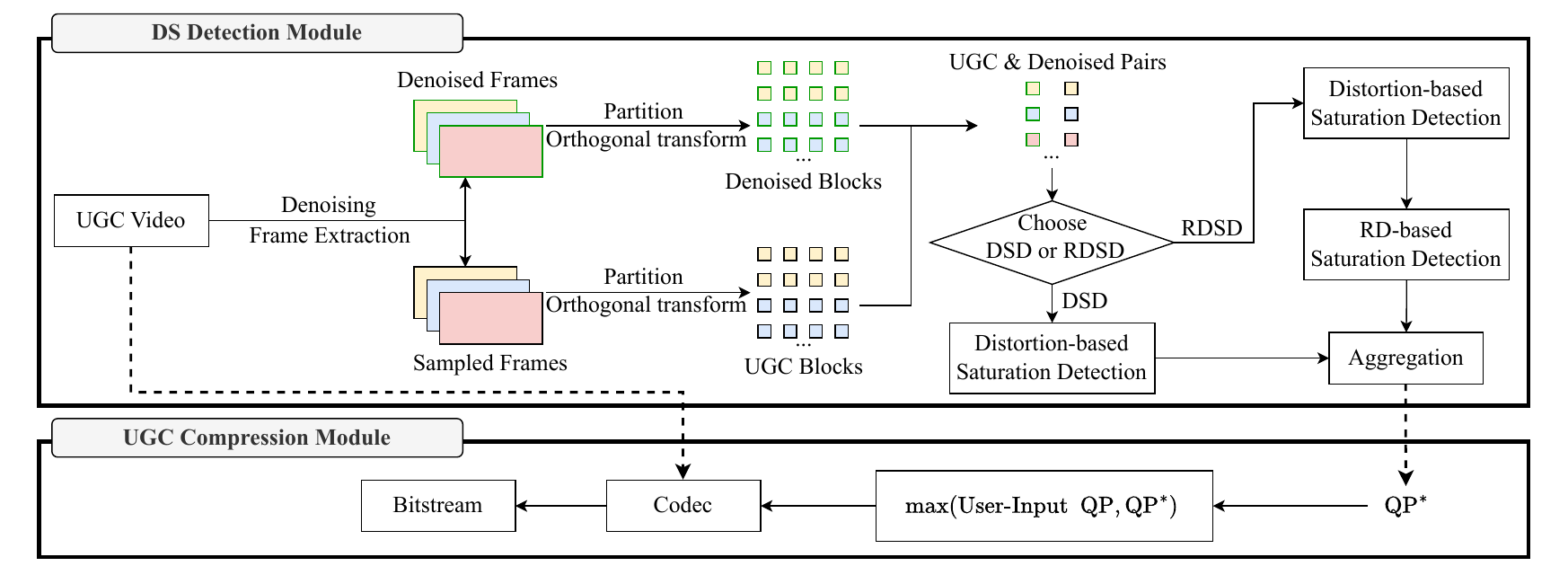}
\caption{The proposed UGC compression pipeline. For UGC input, we first uniformly sample frames and divide the sampled frames into blocks. 
Next, an orthogonal transform is applied to each block.
In this work, we consider only the DCT.
We then detect \dissat{} for each block individually, in terms of a quality parameter. 
Next, we derive a single saturation $\QP^*$ across all blocks. 
This saturation $\QP^*$ is compared with the user-input $\QP$, and the larger one between the two is sent to the codec.}
\label{fig:Method_Pipeline}
\end{figure*}

\section{Motivation}
\label{Method}
We introduce the UGC compression problem from an information-theoretic perspective (\autoref{Method_A}), showing how it relates to \dissat{} \cite{pavez2022compression} and how denoising arises as an important element of UGC compression. We  review the fundamentals of rate-distortion optimization  in \autoref{Method_B}, which we later use  to build our \dissat{} detection methods.

\textbf{Notation.} Uppercase bold letters, such as $\mathbf{U}$, denote random vectors. Lowercase bold letters, such as $\mathbf{x}$, denote deterministic vectors. The $k$th entry of $\mathbf{U}$ is $U_k$ and the $k$th entry of $\mathbf{x}$ is $x_k$. Regular letters denote scalar values. 

\subsection{Noisy Source Coding and Distortion Saturation}
\label{Method_A}
In traditional source coding, the optimal distortion-rate function for coding UGC $\UGC$ with rate $R$ has the form \cite{berger1971rate}
\begin{equation}\label{eq_RD_traditional}
  D(R) =  \mathbb{E}[\Vert \UGC - \hat{\UGC}(R) \Vert^2_2].
\end{equation}
As $R$ increases $\hat{\UGC}(R)$ will approach $\UGC$ and $D(R)$ will decay to zero. However, because $\UGC$ may have low quality (e.g., due to noise), small 
$\Vert \UGC - \hat{\UGC}(R) \Vert^2_2$ does not guarantee that $\hat{\UGC}(R)$ has good perceptual quality \cite{wang2019youtube}. 
Thus, we consider the noisy source coding framework, where the distortion is computed relative to the pristine input 
$\mathbf{X}$ \cite{pavez2022compression}, leading to 
the optimal distortion-rate function \cite{berger1971rate}
\begin{equation}
D^{\text{ugc}}(R) =  \mathbb{E}[\Vert \mathbf{X} - \hat{\UGC}(R) \Vert^2_2]. 
  \label{eq:opt-pristine}
\end{equation}
The noisy source coding theorem \cite{wolf1970transmission} establishes that  \eqref{eq:opt-pristine} can be decomposed as:
\begin{equation}\label{eq_RD_ugc}
  D^{\text{ugc}}(R) = D_0 + \mathbb{E}[\Vert \mathbf{Y} - \hat{\UGC}(R) \Vert^2_2],
\end{equation}
where $D_0 = \mathbb{E}\left[\Vert \mathbf{X} - \mathbf{Y} \Vert\right]^2_2$,  and $\mathbf{Y}$ is the minimum mean squared error estimator (MMSEE) of $\mathbf{X}$ given $\UGC$ \cite{pavez2022compression}.  
Two conclusions follow from \eqref{eq_RD_ugc}.  
First, the optimal UGC encoding can be achieved by optimally denoising  $\UGC$ followed by optimal source coding of the result.
Second, the optimal distortion-rate function for UGC $D^{\text{ugc}}(R)$ saturates as the rate increases. 
More precisely, the minimum achievable distortion when encoding UGC is given by $D_0>0$ and is achieved as the rate $R$ goes to infinity.
Implementing this theoretically optimal UGC compression system 
is impossible since $\mathbf{X}$ and $\mathbf{Y}$ are unavailable. 
Thus, our previous work \cite{pavez2022compression}  proposed using  a traditional codec that optimizes MSE with respect to an alternative reference (\armse):
\begin{equation}
\hat{D}^{\text{ugc}}(R) =  \mathbb{E}[\Vert \mathbf{Z} - \hat{\UGC}(R) \Vert^2_2],
  \label{eq:opt-denoised}
\end{equation}
 where $\mathbf{Z}$ is obtained using an off-the-shelf denoiser to approximate $\mathbf{Y}$, and $\hat{\UGC}(R)$ is 
 obtained by a traditional video codec.
Note that, similar to \eqref{eq:opt-pristine}, the distortion in \eqref{eq:opt-denoised} will saturate to $\mathbb{E}[\Vert \mathbf{Z} - \UGC \Vert^2_2]$ as $R$ increases. 
Although \cite{pavez2022compression} proposed a simple criterion for detecting \dissat{}, it is prone to failure in certain scenarios \cite{xiong2023rate} (see \autoref{Exp}). We propose a more robust and efficient method in see \autoref{Method_B}.

\subsection{Overview of rate-distortion optimization (RDO)}
\label{Method_Overview}
Modern video coding systems rely on rate-distortion optimization (RDO) to choose coding parameters, such as block-level quantization step and block-partitioning \cite{ortega1998rate}, for each input image. 
Let $\img$ be a vectorized image with $NM$ pixels and $\cimg(\thetavec)$ its compressed version using parameters $\thetavec\in\Theta$, where $\Theta\subset\mathbb{N}^{M}$ is the set of all possible operating points and $M$ is the number of blocks. Given blocks of size $N$, $\img_k\in\mathbb{R}^{N}$ for $k = 1, \hdots, \nummb$, we aim to find \cite{ortega1998rate}:
\begin{equation}
    \thetavec^\star = \argmin_{\thetavec \in \Theta} \, \sum_{k = 1}^{\nummb} \, \norm{\img_k - \cimg_k(\pmb{\theta})}_2^2 + \lambda \sum_{k = 1}^{\nummb}\, r_k(\cimg_k(\thetavec)),
\end{equation}
where $r_k(\cdot)$ is the rate for the $k$th coding unit, and $\lambda\geq 0$ is the Lagrange multiplier that controls the RD trade-off. 
When each coding unit can be optimized independently, we obtain $\cimg_k(\thetavec) = \cimg_k(\theta_k)$ \cite{fernandez2025image}, and the problem simplifies to
\begin{equation}
\label{eq:final_form}
  \theta_k^\star = \argmin_{\theta_k \in \Theta_k} \ \norm{\img_k -   \cimg_k(\theta_k)}_2^2 + \lambda \, r_{k}(\cimg_k(\theta_k)),   
\end{equation} 
where $\Theta_k$ is the set of coding parameters for the $k$th block. In modern codecs, the user selects a $\QP$ that controls the quantization step $\quantstep$ and $\lambda$. For instance, a choice for AVC is \cite{wiegand2003rate}: 
\begin{equation}
\label{eq:multiplier_avc}
    \quantstep(\QP) = 2^{(\QP - 4)/6}, \quad \text{and} \quad \lambda = r \, 2^{(\QP-12) / 3},
\end{equation}
where $r$ varies with the type of frame, content, and codec \cite{ringis_disparity_2023}. Next, we introduce our \dissat\ detection methods and explain how they work with RDO-based codecs.

\section{Proposed \dissat\ detection methods}
\subsection{Overview} 
As illustrated in \autoref{fig:Method_Pipeline}, our proposed approach comprises two modules: \dissat{} detection and UGC compression. 
Because modern codecs accept a user‐input \QP\ that jointly controls the quantization step and the Lagrange multiplier (e.g., \eqref{eq:multiplier_avc} for AVC), our \dissat{} detection module only estimates one saturation $\QP^*$ so the target codec uses the maximum of the user-specified \QP\ and $\QP^*$ for coding. This design preserves compatibility with existing video coding pipelines (no modification to the target codec). 

Since \dissat{} depends on properties of the input video that do not change rapidly, we sample the UGC video (e.g., $1$ frame/second) and denoise each sampled frame. 
We then use the UGC and denoised UGC frames to perform block-wise saturation detection in the DCT domain. The detected block-wise saturation results are aggregated to obtain a saturation $\QP^*$, which is provided to the target codec. 

Although our \dissat{} predictors can operate at the block level, and can in principle be integrated into the encoder to prevent \quasat{} block-wise, a \textit{frame-wise saturation $\QP^*$} is calculated at fixed temporal intervals  (e.g., once every second) and kept constant until the following update. 
We choose this approach for three reasons: 
(i) it allows us to maintain compliance with existing video coding pipelines, which typically select a target $\QP$ for each group of pictures and control fine-grain rate adaptation with a Lagrange multiplier $\lambda$ derived from the $\QP$;
%
(ii) Computational efficiency and latency: \dissat{} detection for every frame would be substantially more costly,   
while estimating \dissat{} for frames at fixed intervals yields a lightweight pre-encoding decision that scales to long UGC videos. 
(iii) Stability of perception: per-block saturation prevention may introduce strong spatial \QP\ fluctuations that can induce visual inconsistency; a single clip-level $\QP^*$ provides a robust operating point.
Though block-level integration remains a feasible extension, the proposed clip-level pipeline offers a good trade-off between robustness, complexity, and deployability.

For the \dissat{} detection methods, \OursD\ (see \autoref{Method_B}) finds a saturation quantization step, which is used to determine a saturation $\QP^*_k$ for the $k$th block via \eqref{eq:multiplier_avc}.
\OursRD\ (see \autoref{Method_C}) relies on the output of \OursD: for the $k$th block, based on the saturation point identified by \OursD\, \OursRD\ estimates a saturation Lagrange multiplier $\lambda_{s,k}^{*}$ as the slope at that point using a low-complexity codec.
We then transfer this multiplier to a target codec (e.g., AVC), denoting the transferred multiplier as $\lambda_{t,k}^{*}$, and derive the saturation $\QP^*_k$ from $\lambda_{t,k}^{*}$ via \eqref{eq:multiplier_avc}.  
The final $\QP^*$ is computed as the mean of $\QP^*_k$ (resulting from \OursD{} or \OursRD{}) across the sampled frames.

\begin{figure}[tb]
\centering
\includegraphics[width=1\linewidth]{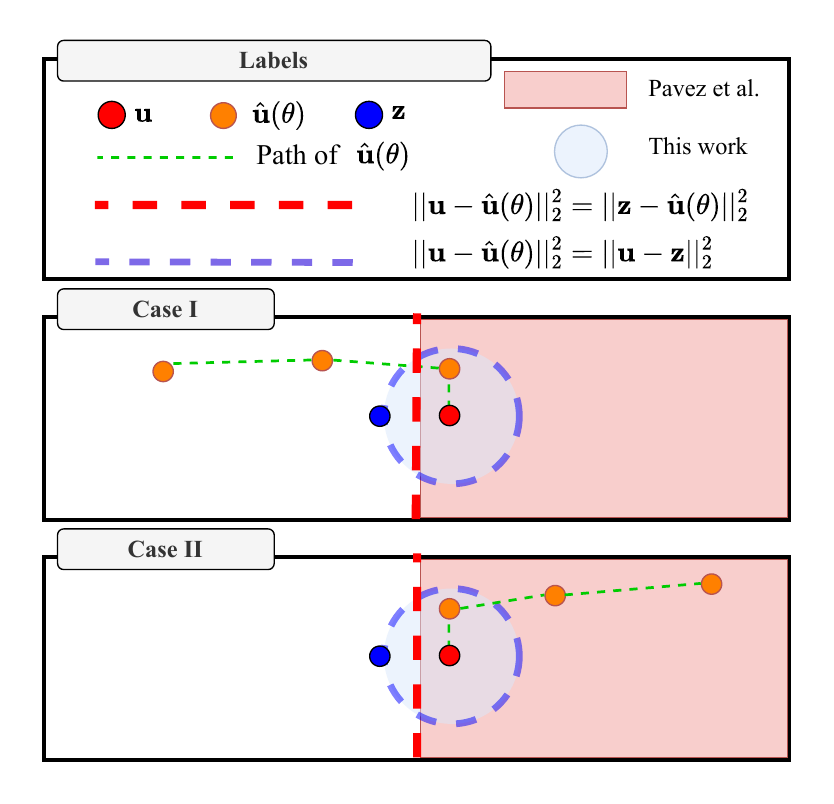}
\caption{
Case I and Case II are possible scenarios for saturation. The UGC input is encoded with different $\theta$ values. The green dotted lines show how the reconstructed signals (orange) approach the UGC input (red) and the denoised UGC (blue). 
The light pink region depicts the saturation region as defined in \cite{pavez2022compression}, whereas the light blue region represents a more robust saturation region defined in \eqref{equ_CostFunc}.}
\label{fig:I4_Sat_Bound}
\end{figure}

\subsection{Distortion-based saturation detection (\OursD) }
\label{Method_B}
Let $\ugc$ and $\dr$ denote the transform-domain representation of a UGC block and its denoised version, respectively. 
$u_{i}$ and $z_{i}$ for $i=1, 2\cdots,N$ represent the $i$th frequency of $\ugc$ and $\dr$, and $\theta$ is the coding parameter for the given block, while $\quantstep_i(\theta)$ is the quantization step for frequency $i$. 

Our prior work \cite{pavez2022compression} concludes that \dissat\ happens when the reconstructed signal is closer to the input UGC than to the denoised reference. 
Thus,  parameters $\theta$  satisfying
\begin{align}
   \Vert \hat{\ugc}(\theta) - \ugc \Vert^2_{2} \leq
  \Vert \hat{\ugc}(\theta) - \dr \Vert^2_{2} ,\label{equ_Method_A_Transform_Domain_Saturation}
\end{align}
should be avoided. \cite{pavez2022compression} defines the saturation point as the intersection of two RD curves, that is,  when $\Vert \hat{\ugc}(\theta) - \ugc \Vert^2_{2} = \Vert \hat{\ugc}(\theta) - \dr \Vert^2_{2}$.
While this criterion captures the intuition behind saturation provided by \eqref{eq:opt-denoised}, it has several problems. 
First, because the saturation region forms a hyperplane (see \autoref{fig:I4_Sat_Bound}), some blocks exhibit saturation (i.e., \eqref{equ_Method_A_Transform_Domain_Saturation} is satisfied) for all values of $\theta$ and consequently other criteria need to be used to choose a saturation parameter.
Second, the quantity  $\Vert \hat{\ugc}(\theta) - \ugc \Vert^2_{2} -  \Vert \hat{\ugc}(\theta) - \dr \Vert^2_{2}$ can be non-monotonic as a function of $\theta$, and its oscillatory behavior makes the saturation point non unique and hard to choose. 
Third, evaluating \eqref{equ_Method_A_Transform_Domain_Saturation} requires quantization and computing distortion at multiple $\theta$, substantially increasing computational complexity.
We make several modifications to \eqref{equ_Method_A_Transform_Domain_Saturation} to address these problems. 

Since the quantized frequency coefficients of a block are usually sparse, we define the following set:
\begin{align}
    \mathcal{T} = 
    \left\{i: \hat{u}_{i}(\theta) = 0, \forall \theta \right\} 
    =\left\{i: \vert u_i \vert < \quantstep_i(\theta)/2, \forall \theta \right\}.
    \label{equ_Method_A_Transform_Domain_theta}
\end{align}
No saturation occurs in these frequencies: the distortion remains constant as the rate changes because they are quantized to $0$ for all $\theta$.
Thus, we exclude them from saturation detection. 
\begin{definition}\label{def_saturation}
We define the saturation region as the set of all coding parameters where the  reconstructed UGC satisfies
\begin{align}
(1/12)\sum_{i\in \mathcal{T}^{c}} \,{\quantstep^2_{i}(\theta)} \leq \sum_{i\in \mathcal{T}^{c}}\,\left | u_{i} -  z_{i} \right | ^ {2} .
  \label{equ_CostFunc}
\end{align}
The saturation point $\theta^*$ satisfies \eqref{equ_CostFunc} with the smallest rate.
\end{definition}
In practice, since  $\theta$ has a one-to-one mapping 
to $q_i(\theta)$, $\theta^{*}$ can be chosen as the smallest $\theta$ satisfying \eqref{equ_CostFunc}.
Our definition of saturation region in \eqref{equ_CostFunc} can be justified by the next result.
\begin{proposition}\label{proposition_saturation}
    If $\theta$ satisfies \eqref{equ_CostFunc} and under high-rate regime  assumptions \cite{gish1968aymptotically}, we have
    \begin{equation}
        \mathbb{E}\left[ \sum_{i\in \mathcal{T}^{c}} \vert \hat{U}_i(\theta) - u_i\vert^2 \right] \leq \mathbb{E}\left[ \sum_{i\in \mathcal{T}^{c}}  \vert \hat{U}_i(\theta) - z_i\vert^2 \right].\label{equ_CostFunc_expectation}
    \end{equation}
    \begin{proof}
        The left-hand sides of \eqref{equ_CostFunc} and \eqref{equ_CostFunc_expectation} are equal at high bitrates, where quantization noise can be accurately modeled as additive, independent, and uniformly distributed \cite{gish1968aymptotically}. By the linearity of expectation and Jensen's inequality,  $\mathbb{E}[\vert \hat{U}_i(\theta) - z_i \vert^2] \geq \vert \mathbb{E}[\hat{U}_{i}(\theta)] -  z_{i}\vert^2 = \vert  u_{i} -  z_{i}\vert^2$. 
    \end{proof}
\end{proposition}
The saturation criteria from \eqref{equ_CostFunc} is related to  \eqref{equ_Method_A_Transform_Domain_Saturation} by \autoref{proposition_saturation} under typical quantization error modeling assumptions \cite{gish1968aymptotically}.
\autoref{fig:I4_Sat_Bound} illustrates the difference in the saturation regions defined by \eqref{equ_CostFunc} and \eqref{equ_Method_A_Transform_Domain_Saturation}. 
Since explicitly visualizing our chosen set of significant frequencies in \eqref{equ_Method_A_Transform_Domain_theta} and the expectation of distortion is not straightforward, for convenience, we present a simpler saturation bound, $ \|\hat{\ugc}(\theta) - \ugc \|_2^2 \leq \norm{\ugc - \dr }_2^2 $. 
This simplified form omits the frequency selection and distortion expectation but is sufficient to illustrate the saturation concept encapsulated in \eqref{equ_CostFunc}.
From \autoref{fig:I4_Sat_Bound}, we observe that $\|\hat{\ugc}(\theta) - \dr\|^2_{2}$ serves as a very loose bound for $\|\hat{\ugc}(\theta) - \ugc\|^2_{2}$, potentially causing certain blocks to appear saturated at extremely low rates. 

In contrast, our proposed criterion \eqref{equ_CostFunc} employs a tighter, fixed bound that depends solely on the intensity of denoising applied to the most significant frequencies.
The underlying rationale in \eqref{equ_CostFunc} is straightforward: since our goal is to avoid allocating bits to noise, if the quantization error is lower than the noise measure, then encoding in that region implies encoding the noise, and we should therefore stop.
Furthermore, by approximating the distortion through its expectation, we eliminate the need to compute $\hat{\ugc}(\theta)$. 
This also ensures that $\quantstep^2_{i}(\theta)/12 - \lvert u_{i} - z_{i} \rvert$ is monotonic with respect to quantization step. 
Compared to \eqref{equ_Method_A_Transform_Domain_Saturation}, the saturation  parameters can  be determined more efficiently and uniquely using \eqref{equ_CostFunc}.

Guided by \eqref{equ_CostFunc}, we focus on the quantization step as the key parameter for saturation. Modern codecs typically use a one-to-one mapping between the quality parameter (\QP) and the quantization step (e.g., \eqref{eq:multiplier_avc}). \OursD\ exploits this mapping to output a saturation \QP\ from the detected saturation quantization step for each block. 
As illustrated in \autoref{fig:I3_System}, for simplicity and compliance with existing video coding pipelines, our saturation detection methods predict a single saturation $\QP^*$\ to interface with a target codec.
Consequently, we here focus on \QP\ as the representative coding parameter, i.e., in the next section, we consider $\theta = \QP$.

\subsection{Rate-distortion-based saturation detection (\OursRD)}
\label{Method_C}
Our \OursD\ method identifies the saturation quantization step for each block, and the final saturation $\QP^*$ reflects the averaged saturation quantization step over all the blocks. However, if we provide this $\QP^*$ to the codec, $\lambda$ is internally derived from \eqref{eq:multiplier_avc}, and could be suboptimal for UGC in practice \cite{ringis2023disparity}.
This discrepancy arises because the proportionality parameter $r$ in \eqref{eq:multiplier_avc} varies with the content; the value commonly used in current codecs is statistically derived from pristine videos and does not reflect UGC characteristics.
Since $\lambda$ sets the trade-off between rate and distortion, our goal here is to directly determine per-block saturation $\lambda^*$ and then find a $\QP^*$ after aggregation to convey this saturation information of Lagrange multiplier to the codec.
We first show how to estimate $\lambda$ at the saturation point for a low-complexity codec used for generating RD points around the saturation. 
Then, we provide results to transfer the estimated $\lambda$ from our low-complexity codec (used to estimate the saturation $\lambda$) to target codecs (used to encode the UGC video).

\subsubsection{Estimating the Lagrange multiplier}
\label{estimate_lambda}
For the $k$th block and its corresponding saturation $\QP_k^*$ (detected by \OursD), we find a saturation $\lambda^{(k)}$ by calculating the approximate slope on the RD curve:
\begin{align}
    S^{(k)}_c = - \frac{\Vert \hat{\ugc}(\QP_k^* + c) - \ugc \Vert^2_2 - \Vert \hat{\ugc}(\QP_k^*) - \ugc \Vert^2_2}
    {R(\hat{\ugc}(\QP_k^* + c)) - R(\hat{\ugc}(\QP_k^*))},
\end{align}
where $R(\hat{\ugc}(\QP_k^*))$ is the rate of encoding the block $\UGC$ with $\QP_k^*$ and $c$ determines the neighbor points on the RD curve to be considered.
The saturation $\lambda^{(k)}$ is then defined as:
\begin{align}
    \lambda^{(k)} = \min_c \left\{S^{(k)}_c : S^{(k)}_c > 0 \right\}.
    \label{equ_Method_C_slope}
\end{align}
Determining all coding parameters, such as block partitioning or quantization step, in video compression is complex and involves in-loop dependencies. 
Thus, we use a low-complexity codec derived from AVC, consisting of one transform (the DCT), uniform quantization with a fixed quantization step for every block, and entropy coding; for simplicity, we rely on Context-Adaptive Variable Length Coding (CAVLC) \cite{richardson2002h}. We use this simplified version of AVC for generating RD points for each block and then calculating the slope.
In practice, real codecs involve numerous coding parameters, leading to RD curves that differ from those generated by our low-complexity codec, and consequently may have a different slope at saturation point. 
In the following, we detail a method to transfer the $\lambda^*$ detected by our low-complexity codec to the target codec.

\subsubsection{Transferring the Lagrange multiplier \texorpdfstring{$\lambda$}{lambda}}
\label{transfer_lambda}
\begin{figure}[!t]
\centering
\includegraphics[width=0.9\linewidth]{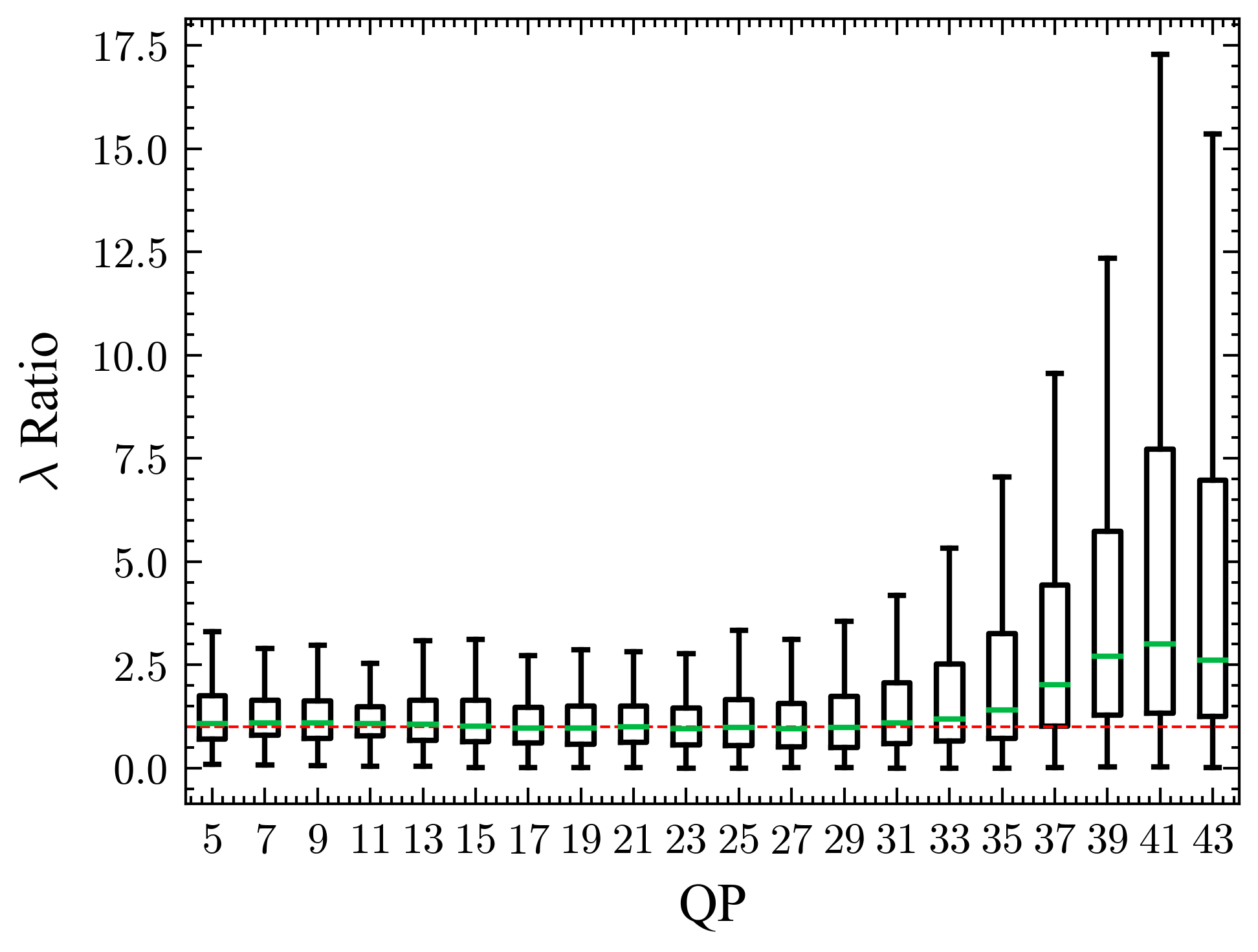}
\caption{Box plot for $a_s / a_t$ in \eqref{equ_Method_D_Equal_DM_5}, using a low-complexity codec and baseline AVC, with images from Kadid-10K across different $\QP$. The red dashed line shows a ratio equal to $1$. Notably, $a_s / a_t$ remains close to 1 in the high-rate region (low $\QP$), where the high-rate assumption is valid. }
\label{fig:Lambda_Validation_2}
\end{figure}

Upon the onset of saturation, we assume the expected quantization errors across different coding systems to be similar.
This assumption is derived from \eqref{equ_CostFunc}, where the right-hand side represents the inherent noise energy of the UGC—an intrinsic characteristic of the content. 
Formally, let the sub-indices $s$ and $t$ represent the low-complexity and the target codec, respectively.
For each $N$-dimensional block, denote $\lams$, $k = 1, \hdots, M$, the Lagrange multiplier at the saturation point estimated using our low-complexity codec and denote $\lamt$ the corresponding Lagrange multiplier in the target codec. 
In the following, expectations are taken with respect to the distribution of the compressed blocks given the input UGC. Equivalent distortion at the onset of saturation implies:
\begin{align}
    \dis = \dit,  \quad k = 1, \hdots, M,
    \label{equ_Method_D_Equal_DM_1}
\end{align}
where $\dis$ and $\dit$ represent the expectation of the distortion for the $k$th UGC block. 
In the high bitrate regime \cite{gish1968aymptotically}, the expected value of the rate can be modeled using \cite{jayant1984digital} $\ris = Na_{s}\log \big(b_{s}/\dis\big)$, and $\rit = Na_{t}\log\big({b_{t}}/{\dit}\big)$, for $k = 1, \hdots, M$,
where $a_{s}$, $b_{s}$, $a_{t}$ and $b_{t}$ are system-dependent parameters that can be statistically determined. 
We can now state the following result.
\begin{proposition}
\label{prop:multiplier}
    In the high bitrate regime,
    \begin{equation}
    \lamt = \lams \,{a_{s}}/{a_{t}}, \quad \text{for} \ k = 1, \hdots, M.
    \end{equation}
\end{proposition}
\begin{proof}
In the high bitrate regime \cite{gish1968aymptotically}, for systems $s$ and $t$:
\begin{align}
    \lams = -\left(\frac{{\rm d}\ris}{{\rm d}\dis}\right)^{-1}, \quad 
    \lamt = -\left(\frac{{\rm d}\rit}{{\rm d}\dit}\right)^{-1}
    \label{equ_Method_D_Equal_DM_3}.
\end{align}
Computing derivatives, $
    {{\rm d}\ris}/{{\rm d}\dis} = - {Na_{s}}/{\dis}$ and $
    {\rm d}\rit/{{\rm d}\dit} = - {Na_{t}}/{\dit}$. Considering \eqref{equ_Method_D_Equal_DM_1} and \eqref{equ_Method_D_Equal_DM_3}:

\begin{align}
    \lamt = {a_{s}}/{a_{t}} \, \lams\label{equ_Method_D_Equal_DM_5},
\end{align}
which concludes the proof.
\end{proof}
\autoref{prop:multiplier} shows that when saturation happens we can translate the Lagrange multiplier at the saturation point to another codec by applying a correction factor that can be statistically determined (\autoref{fig:Lambda_Validation_2}). 
Current codecs convert the input $\QP$ into a Lagrange multiplier via specific equations such as \eqref{eq:multiplier_avc}.
Our \OursRD\ method uses this relationship in the target codec to determine a saturation quality parameter $\QP_k^*$ for the $k$th block based on the saturation $\lambda^{(k)}_t$. 
The final saturation $\QP^*$ is the averaged $\QP_k^*$ across all the blocks.

\begin{figure*}[!t]
\centering
\includegraphics[width=1\linewidth]{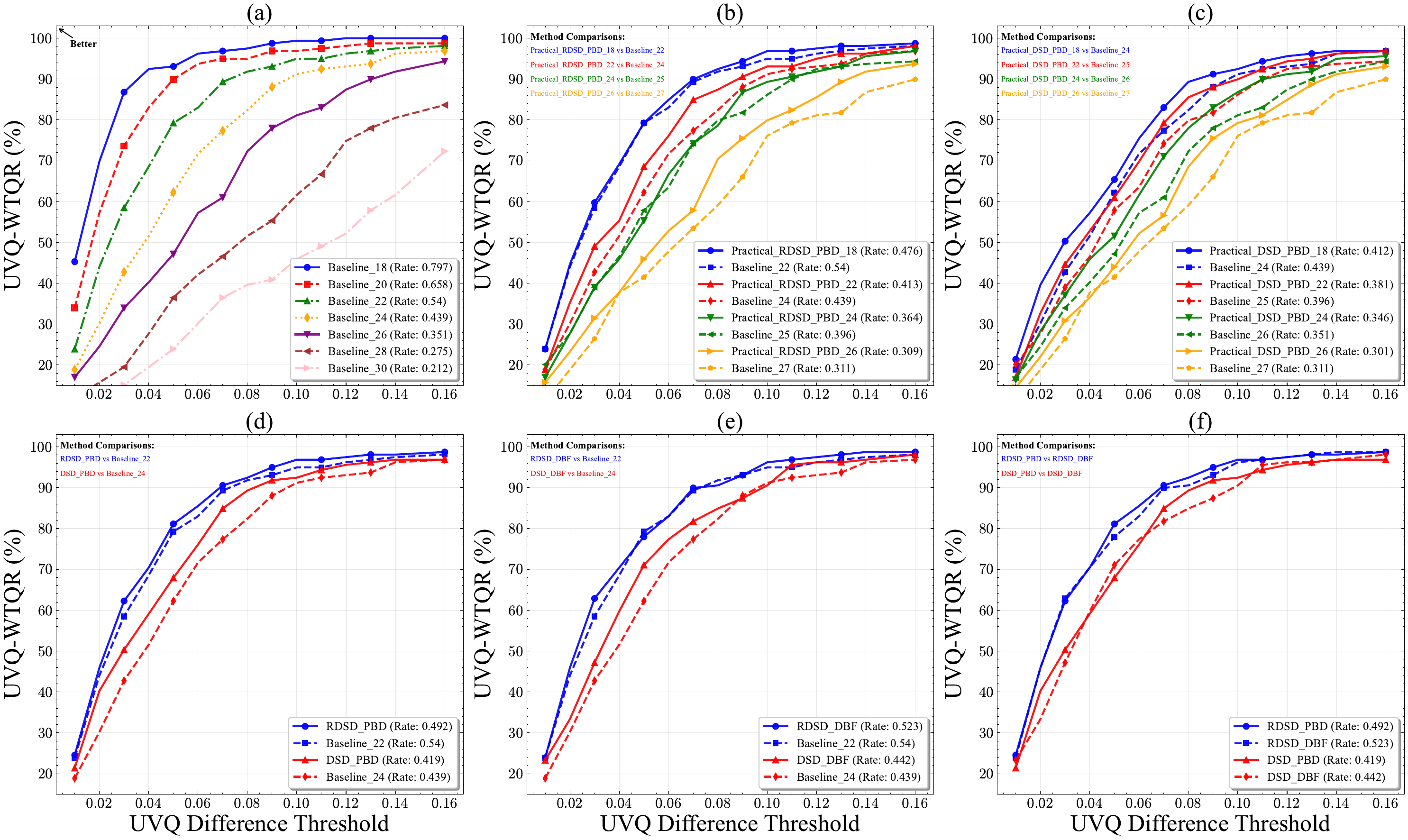}
\caption{
UVQ difference threshold versus the percentage of videos whose UVQ scores decrease by no more than the threshold. 
For method name reference (See~\ref{EXP_Setup}): (a) Baseline. (b–c) Baseline vs. our UGC compression system with \dissat\ detector RDSD and DSD, respectively. (d–e) Baseline vs. encoding at the saturation point predicted by our \dissat\ detectors, using PBD and DBF as denoisers, respectively. (f) Comparison between the two denoisers: PBD and DBF.
}
\label{fig:Threshold_vs_Percentage}
\end{figure*}

\section{Experimental Results}
 \label{Exp}
\subsection{Experimental Setups}
\label{Exp_UGC}
\subsubsection{UGC dataset}
We validate our methods on the YouTube-UGC dataset~\cite{wang2019youtube}, a large-scale public benchmark for UGC video compression and quality assessment, which contains 1,500 clips, each 20 seconds long, spanning 15 categories. Each clip is annotated with a mean opinion score (MOS) obtained from human subject evaluations.
For our experiments, we focus on a subset of the YouTube-UGC dataset~\cite{youtube_media} (link labeled “DMOS”), which includes 159  clips with different quality levels (MOS ranges from 2.5 to 4.5) and three content categories: 61 for Gaming, 60 for Sports, and 38 for Vlog.

\subsubsection{Denoiser for UGC}
\label{Exp_Denoiser}
The primary challenge in UGC denoising is the unpredictability of noise type and intensity. 
Traditional filtering-based denoising methods \cite{dabov2007image, maggioni2012video} require hyperparameter tuning to handle diverse noise characteristics, which can be challenging to set for large UGC datasets. Learning-based denoising methods \cite{zhang2017beyond, chen2018image, zhang2023practical, liang2022vrt} are adaptive to noise levels \cite{zhang2017beyond} and noise types \cite{zhang2023practical}, making them more suitable for UGC denoising.

We use two representative denoisers: Practical Blind Denoiser (PBD) \cite{zhang2023practical} and De-Blocking Filter (DBF) from FFmpeg \cite{tomar2006converting}. We utilize FFmpeg's De-Blocking implementation in post-processing mode 'spp=4:10'. 
PBD is a state-of-the-art deep blind image denoiser designed to address compression artifacts. It is trained on a dataset encompassing multiple noise types and intensities. 
As an extensively trained deep learning–based method, PBD outperforms DBF on most UGC content. Nevertheless, learning-based denoisers are not flawless \cite{ blau2019rethinking}. As illustrated in \autoref{fig:Exp_Denoiser}, both denoisers can excessively smooth areas with rich textures, such as grasslands and nets. Additionally, differences between PBD and DBF are evident in the zoomed-in sections of \autoref{fig:Exp_Denoiser}: PBD is a more aggressive approach, altering many pixels, whereas DBF is more conservative. 
Note that our methods do not depend on a particular denoiser; our experiments show that both PBD and DBF provide good results, despite their differences.

\subsubsection{Quality assessment for UGC}
\label{Exp_Quality}
To validate our methods, we evaluate the quality of UGC clips before and after compression using non-reference metrics (NRMs) \cite{mittal2012making, mittal2012no, wu2023dover, wu2023exploring, sun2022deep, tu2021ugc, wang2021rich, tu2021rapique}. Specifically, we selected four widely used state-of-the-art UGC quality assessment metrics: UVQ \cite{wang2021rich}, SimpleVQA \cite{sun2022deep}, DOVER \cite{wu2023dover} and FasterVQA \cite{wu2022fasterquality}. 
Unlike single-score predictors, UVQ estimates quality across multiple dimensions (semantic content, technical quality, and compression level), while DOVER estimates the quality from both technical and aesthetic perspectives.  In our experiments with UVQ and DOVER, we only use their technical scores,  which are trained to capture the perceptual impact of compression.

\begin{figure}[!t]
\centering
\includegraphics[width=1\linewidth]{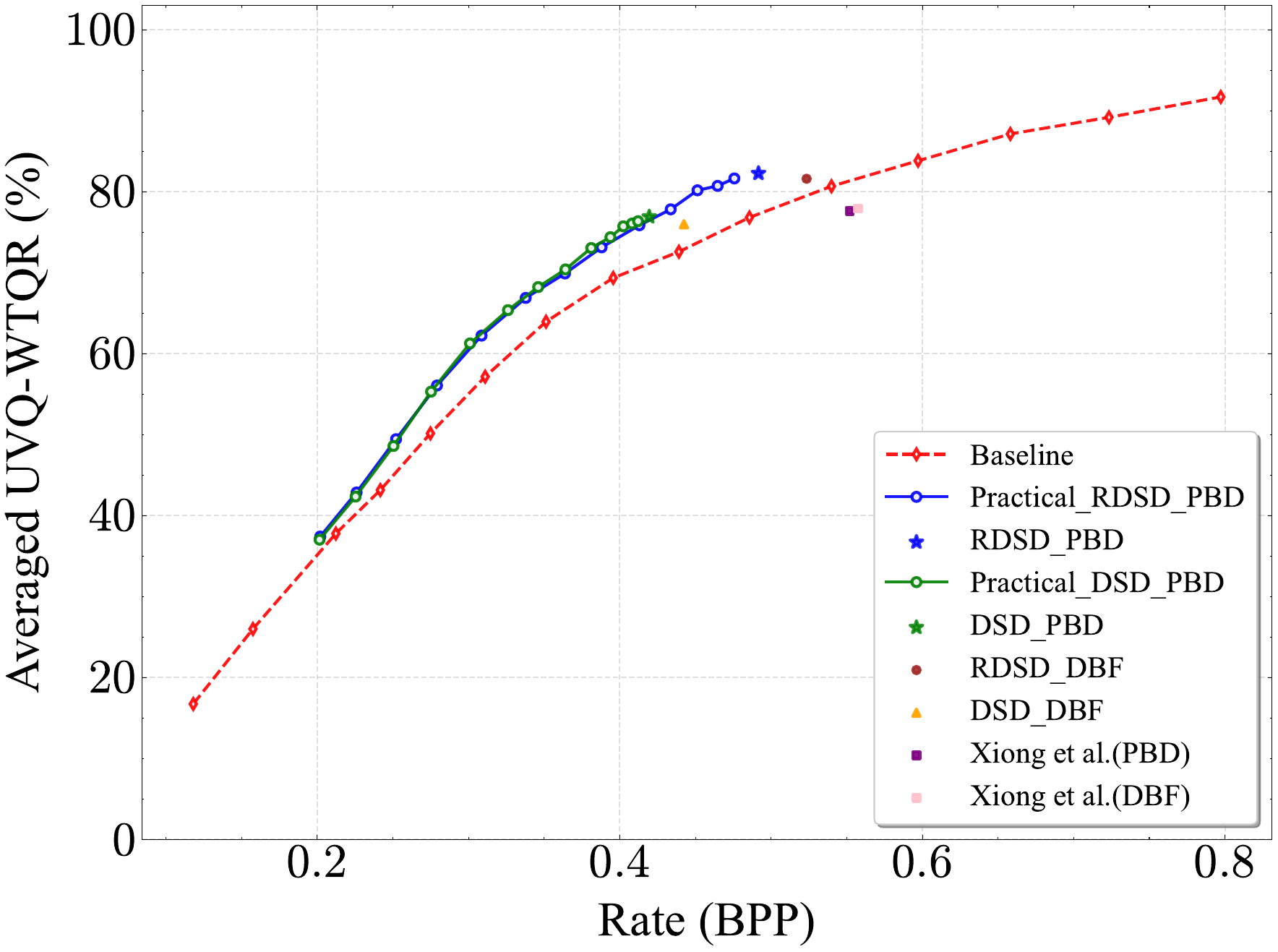}
\caption{
Curve of rate and averaged UVQ-WTQR, which is computed as the mean UVQ-WTQR over all selected thresholds. Solid markers denote RDSD, DSD, and method of~\cite{xiong2023rate}; hollow markers denote the UGC compression system with \dissat\ detectors RDSD and DSD. The red curve indicates the baseline.
}
\label{fig:Rate_vs_AvgPercentage}
\end{figure}

\subsection{Encoding setup}
\label{EXP_Setup}

As illustrated in \autoref{fig:I3_System}, our UGC compression system employs a standard-compliant codec to encode UGC. In this experiment, we use an 4:2:0 AVC encoder (baseline profile) in I–P–P… configuration. The group of pictures (GOP) size is set to 30 frames (1 second). For each GOP, the \QP\ input to the codec is mapped to the Lagrange multiplier $\lambda$ used in RDO to select coding parameters such as block partitioning and the quantization step. The chroma channels are encoded at the same \QP\ as the luma channel. 

In our \dissat{} detection methods, a low-complexity codec is used to generate RD points. This codec is derived from AVC, comprising $4 \times 4$ DCT \cite{strang1999discrete, malvar2003low}, quantization, and Context-Adaptive Variable Length Coding (CAVLC) ~\cite{marpe2003context}.
%
As a baseline, all GOPs are encoded with the same $\QP$ chosen from $\QP \in \{18,  \ldots, 34\}$. For example, in 'Baseline\_18'  all GOPs are encoded with $\QP=18$. 

For each GOP (30 frames), the \dissat\ detectors \OursD, \OursRD, as well as the method in~\cite{xiong2023rate} first sample the middle frame, predict a saturation $\QP^*$ from that frame, and then use the $\QP^*$ to encode the whole GOP. 
We consider two denoisers (PBD and DBF); for notation, 'RDSD\_PBD' indicates that the $\QP^*$ is determined by \OursRD\ using PBD as denoiser.
%
Ideally, we should select the $\QP$ determined by \dissat\, i.e., the best quality that does not result in quality saturation. 
In practice, however, bandwidth budgets may require operating in the non-saturation regime. Consequently, our UGC compression system encodes each GOP at a user-input $\QP \in \{18,  \ldots, 30\}$, if  $\QP>\QP^*$ (see \autoref{fig:I3_System}). Results for this system are prefixed with 'Practical\_'.

\subsection{Setup of \dissat\ detection}
\label{Exp_Setup}
\subsubsection{Setup of \OursD\ and \OursRD}
For each UGC clip, 
we detect \dissat\ using only the luminance channel of each sampled frame and its denoised version, because the human visual system is more sensitive to luminance than to chroma variations \cite{van2001vision}. We detect \dissat\ for each $16 \times 16$ block individually, and the 
frequency coefficients are obtained by applying DCT on each $4 \times 4$ sub-block.
In \OursD\, for the $k$th block, we first detect a saturation quantization step using \eqref{equ_CostFunc} and then apply  \eqref{eq:multiplier_avc} to obtain the saturation $\QP^*_k$. 
In \OursRD\, we estimate the saturation Lagrange multiplier $\lambda^{(k)}_s$ for the $k$th block using \eqref{equ_Method_C_slope} with $c = 5$. 
We then obtain $\lambda^{(k)}_t$ by transferring $\lambda^{(k)}_s$ from our low-complexity codec to the target codec via \eqref{equ_Method_D_Equal_DM_5}.
Next, using \eqref{eq:multiplier_avc} with $r = 0.85$, we derive a saturation $\QP^*_k$ from $\lambda^{(k)}_t$. 
Both \OursD\ and \OursRD\ average $\text{QP}^*_k$ across blocks to obtain a single $\QP^*$ per sampled frame.

\begin{figure}[!t]
\centering
\includegraphics[width=\linewidth]{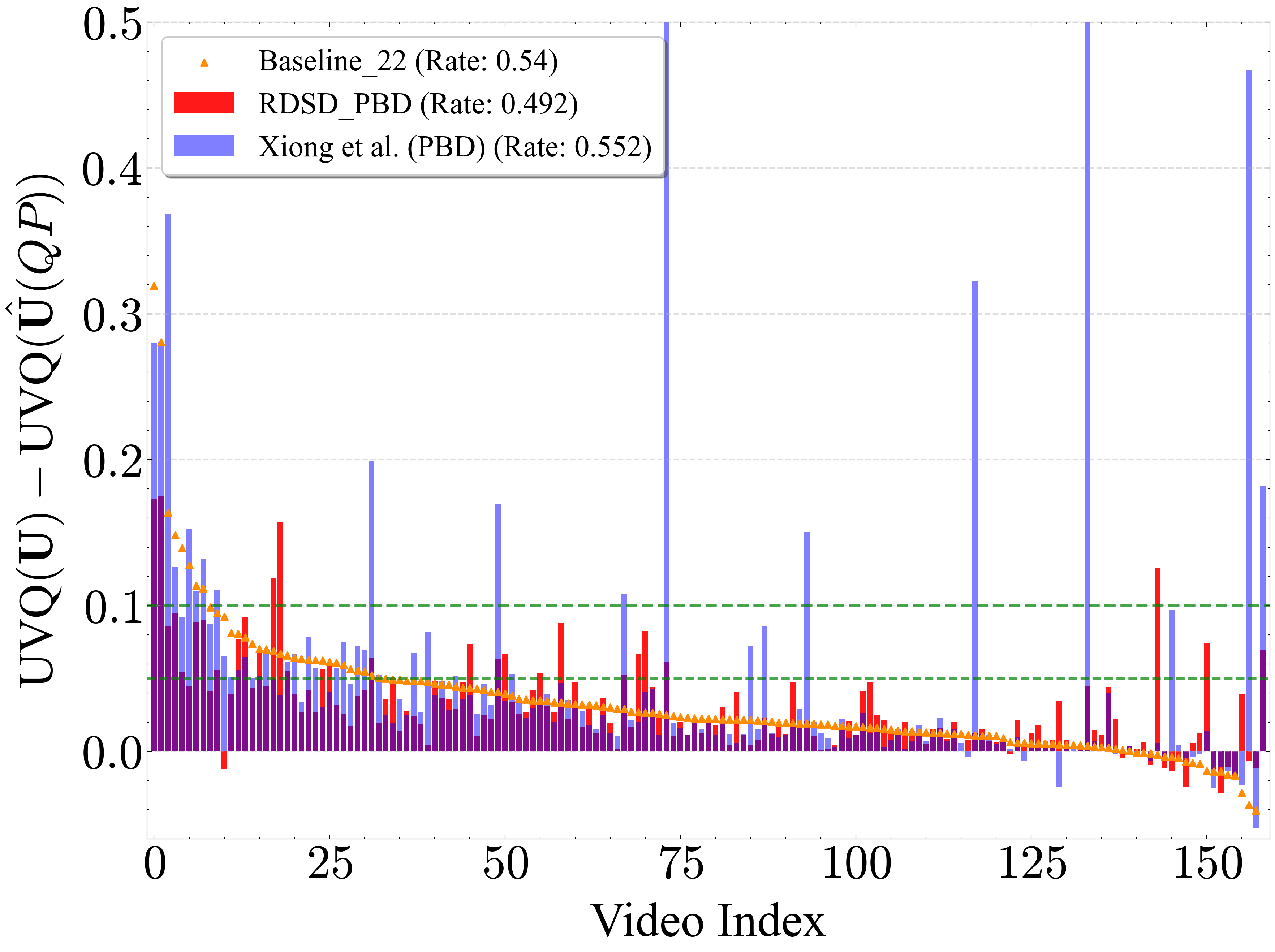}
\caption{Comparison between \OursRD\ and \cite{xiong2023rate} using PBD denoiser. Orange triangles represent the baseline with $\QP = 22$. Each bar represents a UGC clip, with the height of the bar indicating the UVQ score difference after compression. The green lines indicate UVQ score differences of 0.1 and 0.05.}
\label{fig:Comparison}
\end{figure}

\subsubsection{Transferring the Lagrange multiplier \texorpdfstring{$\lambda$}{lambda}}
We determine the ratio \texorpdfstring{$a_{s}/a_{t}$}{as/at} in \eqref{equ_Method_D_Equal_DM_5} empirically. We consider $81$ images from the Kadid-10K dataset \cite{kadid10k} and  encode the luminance channel with both our low-complexity codec and the target codec using a range of quantization steps corresponding to $\QP = 0, \hdots, 45$ in AVC \cite{richardson2002h}. 
To ensure both our low-complexity codec and the target codec encode at the same quantization step (the condition for \eqref{equ_Method_D_Equal_DM_5}), the  target codec is set to encode in constant $\QP$ mode, where the quantization step is fixed for each block.
Then, for each $16 \times 16$ block, we estimate $\lambda_s$ and $\lambda_t$ for each chosen $\QP$ (corresponding to a unique quantization step) by finding the slope on the RD curves (similar to \eqref{equ_Method_C_slope} with $c = 5$). 
We find the ratio $a_{s}/a_{t}$ in \eqref{equ_Method_D_Equal_DM_5} from the statistical distribution of $\lambda_t/\lambda_s$ across all blocks. 

The box chart in \autoref{fig:Lambda_Validation_2} shows the distribution of $\lambda_t/\lambda_s$ across the blocks, with each bar corresponding to a unique quantization step.
We take the median of each distribution as the approximated $a_{s}/a_{t}$ ratio.
In the high-rate region, the ratio is close to $1$, suggesting that the same quantization step (also same distortion under high-rate assumption) results in the same $\lambda$, as predicted by \autoref{prop:multiplier}. However, in the low-rate region, the ratio exceeds $1$ as the high-rate assumption no longer holds. 
Since this ratio varies across quantization steps, for the $k$th block, we use the ratio that corresponds to the $\QP^*_k$ detected by \OursD\ for transferring $\lambda$.


We validate our methods on AVC (baseline profile) as the target codec. We choose AVC because it remains the most widely used codec across all types of services \cite{moina2024cloud}. Nevertheless, this approach can be extended to other compression systems using our method to transfer the Lagrange multiplier between codecs (see \autoref{fig:Lambda_Validation_2}).

\begin{figure*}[!t]
\centering
\includegraphics[width=1\linewidth]{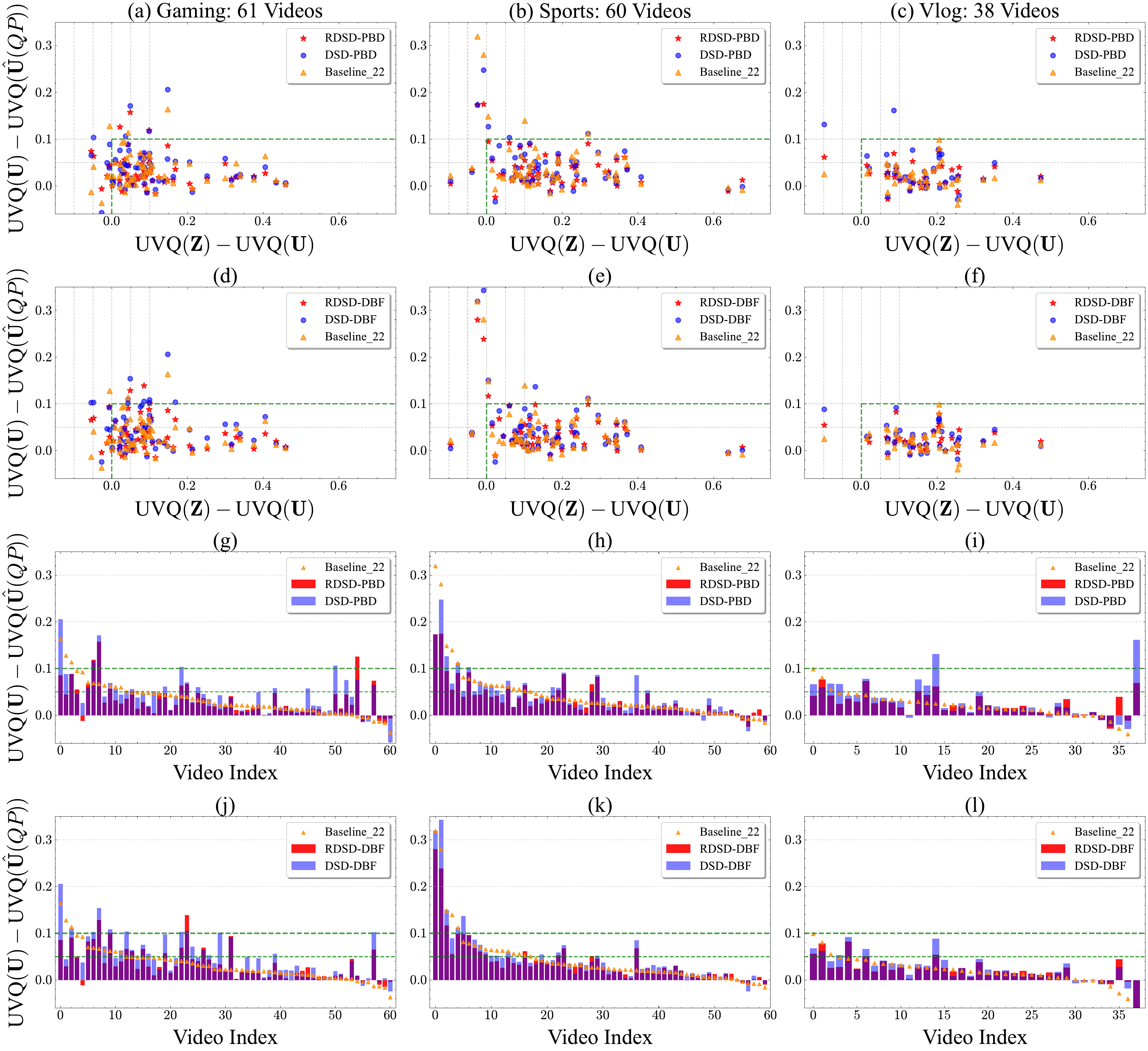}
\caption{Results of encoding at the \dissat{} detected by our methods.
In the scatter plots, each point represents a compressed UGC clip: orange triangles denote clips encoded with the baseline with $\QP=22$, while blue circles and red stars indicate clips encoded at the saturation detected by \OursRD\ and \OursD{}. 
In the bar charts, each bar represents a UGC clip, with its height indicating the UVQ score difference between the original and compressed clips.  
}
\label{fig:Barchat_Scatter}
\end{figure*}

\begin{figure*}[!t]
\centering
\includegraphics[width=\linewidth]{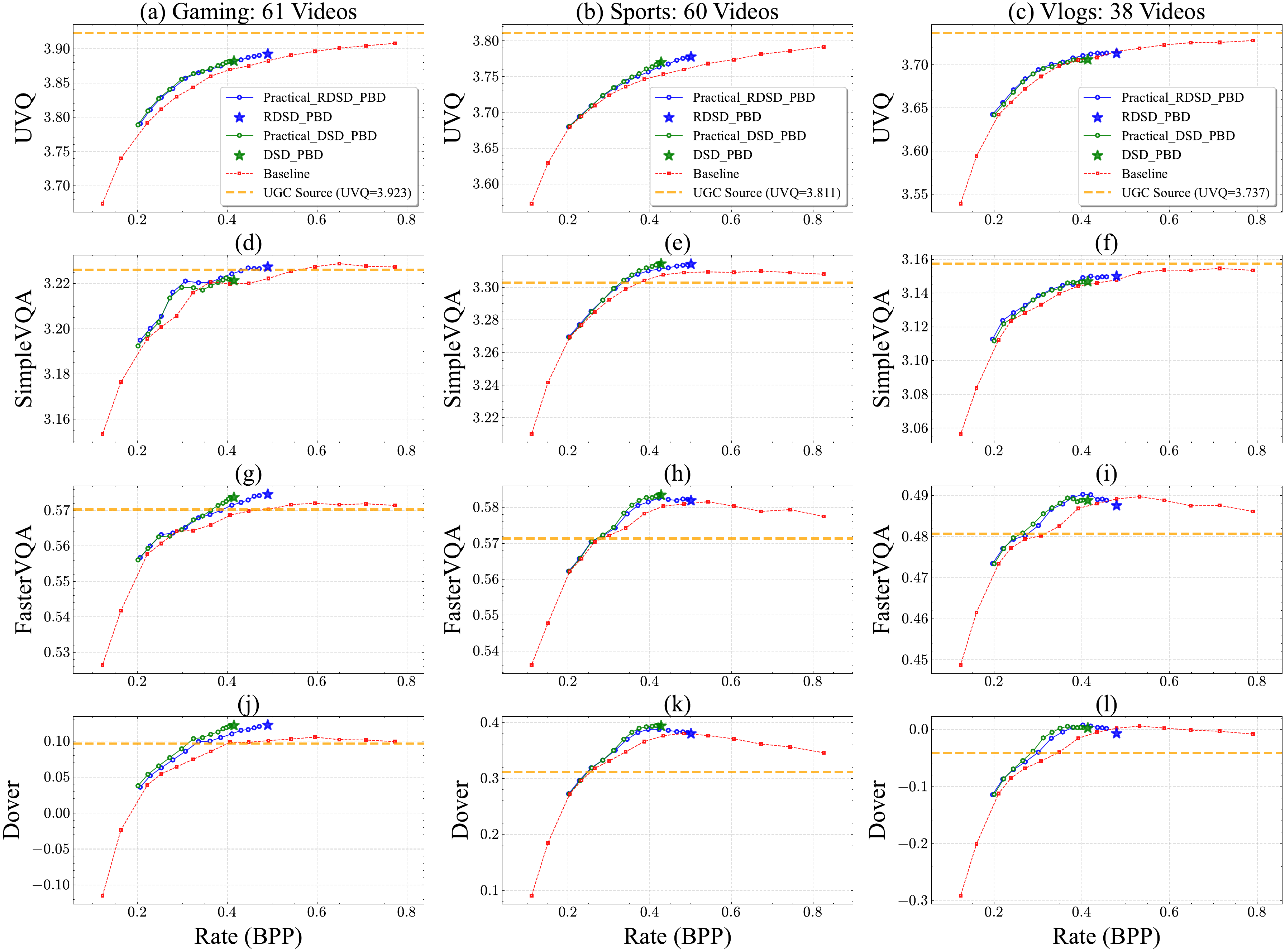}
\caption{
Rate and average scores measured by four NRMs. In each subfigure, the yellow dashed line indicates the scores of the UGC sources, the red dashed line indicates the baseline, solid star markers indicate results obtained with our \dissat\ detectors (RDSD and DSD using PBD as the denoiser), and hollow circle markers indicate the UGC compression systems.
}
\label{fig:Rate_vs_Metrics}
\end{figure*}

\subsection{Results analysis}
\label{Exp_D}
To show that our \dissat\ detection method can help avoid \quasat\ measured by NRMs, 
we present: (i) a threshold-based analysis of quality score degradation, in which the compressed videos are deemed perceptually equivalent to the UGC source when the non-reference score degradation falls below a threshold; and (ii) BD-rate savings of our method against the baseline across multiple NRMs. Finally, we provide visual examples to illustrate the effectiveness of our approach.
We emphasize again that \textit{the NRMs we use to assess performance were not part of the \dissat{} design}.  

\subsubsection{Threshold-based analysis of quality score degradation}
As a way to monitor the quality of user experience, service providers can set a pre-defined threshold on a NRM, which indicates the maximum tolerable quality degradation. 
These thresholds are metric-dependent and typically require extensive subjective studies to establish.
In this experiment, we adopt UVQ as the target NRM for two reasons. First, it includes a branch specifically designed for predicting quality in terms of compression intensity.
Second, its authors provide concrete recommendations: a 0.05 score difference threshold for high-quality clips and a 0.10 threshold for lower-quality clips. 
Since the exact thresholds can vary across clips, and our method is not tailored to any particular thresholds, 
we evaluated a sweep of thresholds from 0.01 to 0.16, encompassing and extending the recommended range. 

\autoref{fig:Threshold_vs_Percentage} reports the UVQ within threshold quality ratio (UVQ-WTQR), defined as the percentage of clips whose UVQ score degradation does not exceed a given threshold.
\autoref{fig:Threshold_vs_Percentage} (a) shows the results of the baseline; as expected, higher bitrate yields a higher percentage at each threshold.
\autoref{fig:Threshold_vs_Percentage} (b) and \autoref{fig:Threshold_vs_Percentage} (c) compare the baseline with our UGC compression system using \OursRD\ and \OursD\ as the \dissat\ predictors, respectively. Compared to the baseline, our UGC compression system attains equal or higher UVQ-WTQR at lower bitrates, indicating bitrate savings at the same quality degradation.
\autoref{fig:Threshold_vs_Percentage} (d) and \autoref{fig:Threshold_vs_Percentage} (e) compare the baseline with encoding at the $\QP^*$ estimated by \dissat\ predictors (\OursRD\ and \OursD) using PBD and DBF as the denoisers, respectively; encoding at the predicted \dissat\ reduces bitrate and our methods work regardless of the choice of denoiser.
Finally, \autoref{fig:Threshold_vs_Percentage} (f) compares the two denoisers: PBD generally outperforms DBF. Because \dissat\ detection relies on noise estimation, better denoising improves \dissat\ detection accuracy; nevertheless, our methods are agnostic to the denoiser choice.

\autoref{fig:Rate_vs_AvgPercentage} presents the rate and averaged UVQ-WTQR trade-off curves. The average percentage is computed as the mean of the percentages across all selected thresholds. At the same average percentage (i.e., comparable compressed-video quality), our methods achieve a lower bitrate, demonstrating their effectiveness. Furthermore, \cite{xiong2023rate} performs worse than both our methods and the baseline. As illustrated in \autoref{fig:Comparison}, the method in \cite{xiong2023rate} fails to detect saturation in some cases, resulting in very large UVQ score differences between the input and the compressed clips. 

The scatter plots and bar charts in \autoref{fig:Barchat_Scatter} show the results of encoding UGC clips at \dissat\ predicted by \OursRD\ and \OursD\ using two denoisers. Each scatter plot compares the UVQ improvement attained by denoising UGC against the UVQ reduction after compression. The bar charts illustrate the magnitude of this UVQ reduction.
We see encoding at the \dissat\ predicted by \OursRD\ tends to have smaller UVQ score degradation compared to \OursD. While our method can work with any off-the-shelf denoiser, different denoisers can yield different performance; for instance, compared to PBD, using DBF as the denoiser results in a greater quality reduction, indicating that PBD is a better choice in terms of coding performance. 

\subsubsection{Rate and quality scores curves}
\label{Exp_Results_Rate_Quality}
Another way to demonstrate that our \dissat\ detection methods help avoid \quasat, as assessed by NRMs, is to examine the trade-off between rate and quality scores. In these experiments, we consider four popular NRMs: UVQ~\cite{wang2021rich}, SimpleVQA~\cite{sun2022deep}, DOVER~\cite{wu2023dover}, and FasterVQA~\cite{wu2022fasterquality}. \autoref{fig:Rate_vs_Metrics} depicts bitrate versus quality scores for each metric; at the same quality scores, our methods consistently achieve lower bitrate than the baseline. The results from \autoref{fig:Rate_vs_Metrics} are summarized in \autoref{tab:BD_Savings}  as  BD-rate savings with respect to the baseline.

\begin{table}[t]
  \centering
  \caption{BD-rate savings (\%) of our UGC compression system using \OursRD\ and \OursD\ as the \dissat\ detector against the baseline. Our methods yield better compression quality as measured by different NRMs.}
  \label{tab:BD_Savings}
  \tabcolsep=3.5pt
  \begin{tabular}{@{} l c c c c c c c c @{}}
    \toprule
    & \multicolumn{2}{c}{\textbf{UVQ}}
    & \multicolumn{2}{c}{\textbf{SimpleVQA}}
    & \multicolumn{2}{c}{\textbf{FasterVQA}}
    & \multicolumn{2}{c}{\textbf{DOVER}} \\
    \cmidrule(lr){2-3}\cmidrule(lr){4-5}\cmidrule(lr){6-7}\cmidrule(lr){8-9}

      & \textbf{RDSD} & \textbf{DSD}
      & \textbf{RDSD} & \textbf{DSD}
      & \textbf{RDSD} & \textbf{DSD}
      & \textbf{RDSD} & \textbf{DSD} \\
    \midrule
    \textbf{Game}  & $-10.8$ & $-11.6$ & $-13.1$ & $-7.2$  & $-12.4$ & $-12.7$ & $-16.5$ & $-21.2$ \\
    \textbf{Sport} &  $-7.7$ &  $-7.7$ & $-9.4$ & $-9.5$ & $-16.8$ & $-18.4$ & $-22.3$ & $-24.1$ \\
    \textbf{Vlog}  &  $-6.4$ &  $-3.7$ & $-9.4$ & $-5.9$ & $-18.7$ & $-22.3$ & $-14.6$ & $-18.6$ \\
    \midrule
    \textbf{All}   &  $-8.5$ &  $-8.3$ & $-10.8$ & $-7.8$ & $-15.7$ & $-16.9$ & $-18.8$ & $-21.1$ \\
    \bottomrule
  \end{tabular}
\end{table}

\begin{figure*}[!t]
\centering
\includegraphics[width=1\linewidth]{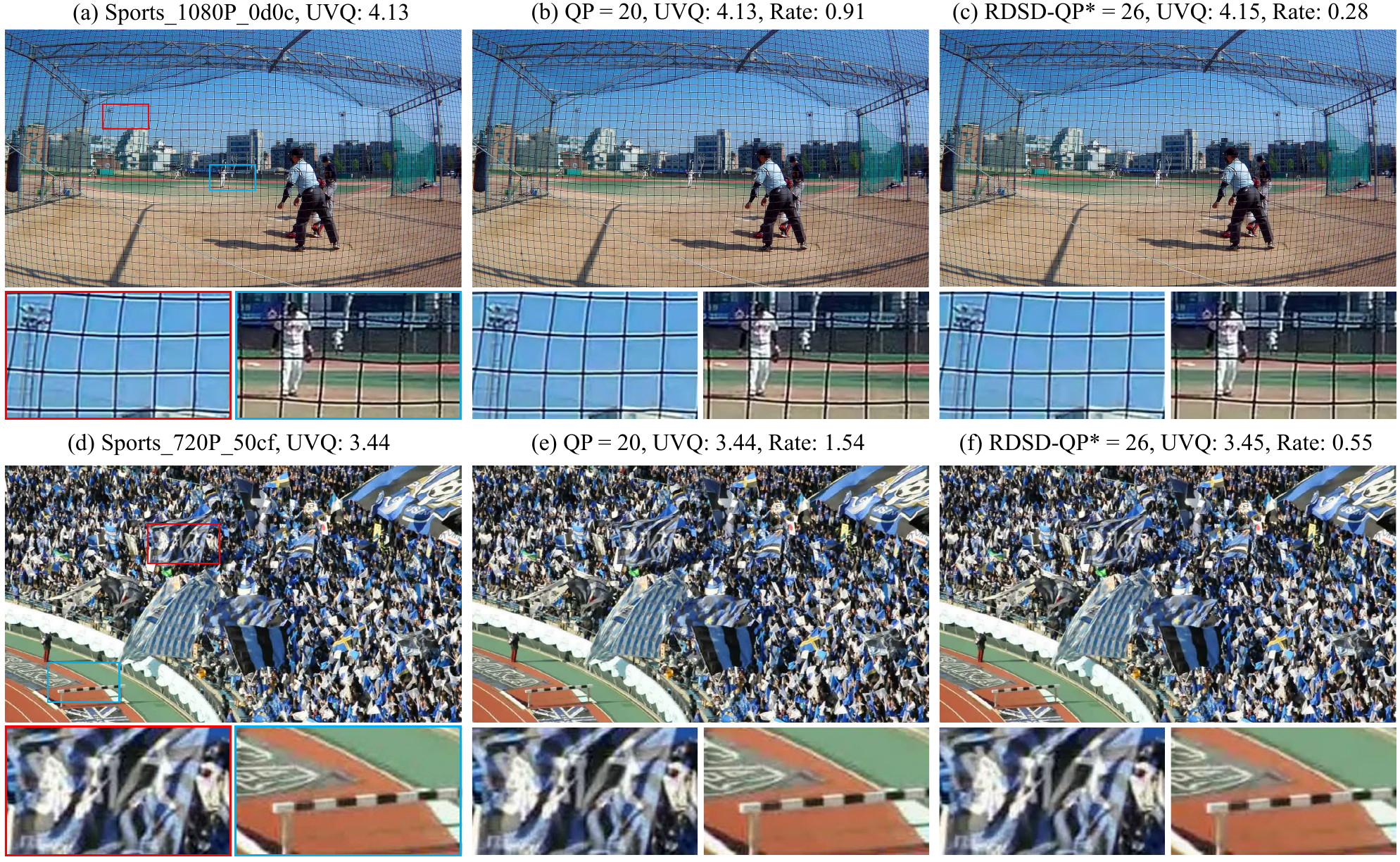}
\caption{From left to right: original UGC, and UGC encoded at $\QP=20$ and the $\QP^*$ given by RDSD. Bitrates are measured in bits per pixel. Our method saves a significant bitrate while producing frames visually identical to the original UGC.
}
\label{fig:Exp_Visualization}
\end{figure*}

\subsection{Visualization examples}
\label{EXP_Visualization}
In \autoref{fig:Exp_Visualization}, we show the results of encoding two UGC clips, where our saturation detection method identifies $\QP^* = 26$. The figures reveal minimal visual differences between the original UGC frames and those encoded at either $\QP = 20$ or $\QP^*$. This indicates that encoding with $\QP^*$ does not visibly degrade quality while offering a much lower bitrate than encoding with $\QP = 20$. These examples illustrate the efficiency of our methods in avoiding \quasat\ for UGC compression.

\section{Conclusion}
\label{Conclusion}
In this work, we tackle UGC compression by first detecting \dissat\ with denoised references and then avoiding \quasat\ using our UGC compression system.
We introduced two \dissat\ detection methods: \OursD, which relies on an input-dependent threshold derived from the \armse\ of the input UGC, and \OursRD, which estimates the Lagrangian at the saturation point.
Experiments on the YouTube UGC dataset show that both methods effectively prevent \quasat\ and achieve substantial bitrate savings.
Moreover, empirical results demonstrate that larger bitrate savings are achievable with more precise \quasat\ avoidance.

As future work, we are interested in exploring denoising strategies that can better adapt to a target NRM \cite{sfm2025rdo, blau2019rethinking} and to different types of UGC \cite{wang2019youtube}. 
Furthermore, both the theoretical relationship between I-MSE, D-MSE, and P-MSE \cite{man2025proxies} and the connection between \dissat\ and \quasat\, are worth exploring. 
Additionally, we aim to integrate \quasat\ as an additional constraint in the RDO process and adapt RDO to directly optimize NRMs during encoding \cite{sfm2025rdo}.

\bibliographystyle{IEEEtran}
\bibliography{refs}


 





\end{document}